\documentclass[12pt,preprint]{emulateapj}

\shorttitle{$z \sim 2$ Emission-Line Galaxies}
\shortauthors{Trump et al.}

\begin{document}

%\slugcomment{\bf Draft: \today}

\title{A CANDELS WFC3 Grism Study of Emission-Line Galaxies at $z \sim
  2$: A Mix of Nuclear Activity and Low-Metallicity Star
  Formation\altaffilmark{*}}

\author{
  Jonathan R. Trump,\altaffilmark{1}
  Benjamin J. Weiner,\altaffilmark{2}
  Claudia Scarlata,\altaffilmark{3}
  Dale D. Kocevski,\altaffilmark{1}
  Eric F. Bell,\altaffilmark{4}
  Elizabeth J. McGrath,\altaffilmark{1}
  David C. Koo,\altaffilmark{1}
  S. M. Faber,\altaffilmark{1}
  Elise S. Laird,\altaffilmark{5}
  Mark Mozena,\altaffilmark{1}
  Cyprian Rangel,\altaffilmark{5}
  Renbin Yan,\altaffilmark{6}
  Hassen Yesuf,\altaffilmark{1}
  Hakim Atek,\altaffilmark{7}
  Mark Dickinson,\altaffilmark{8}
  Jennifer L. Donley,\altaffilmark{9}
  James S. Dunlop,\altaffilmark{10}
  Henry C. Ferguson\altaffilmark{9}
  Steven L. Finkelstein,\altaffilmark{11}
  Norman A. Grogin,\altaffilmark{9}
  Nimish P. Hathi,\altaffilmark{12}
  St\'{e}phanie Juneau,\altaffilmark{2}
  Jeyhan S. Kartaltepe,\altaffilmark{8}
  Anton M. Koekemoer,\altaffilmark{9}
  Kirpal Nandra,\altaffilmark{13}
  Jeffrey A. Newman,\altaffilmark{14}
  Steven A. Rodney,\altaffilmark{15}
  Amber N. Straughn,\altaffilmark{16}
  and Harry I. Teplitz\altaffilmark{7}
}

\altaffiltext{*}{
  Based on observations with the NASA/ESA \emph{Hubble Space
  Telescope}, obtained at the Space Telescope Science Institute, which
  is operated by AURA Inc, under NASA contract NAS 5-26555.
\label{candels}}

\altaffiltext{1}{
  University of California Observatories/Lick Observatory and
  Department of Astronomy and Astrophysics, University of California,
  Santa Cruz, CA 95064 USA
\label{UCO/Lick}}

\altaffiltext{2}{
  Steward Observatory, University of Arizona, 933 North Cherry Avenue,
  Tucson, AZ 85721 USA
\label{Arizona}}

\altaffiltext{3}{
  Astronomy Department, University of Minnesota, Minneapolis, MN 55455
  USA
\label{Minnesota}}

\altaffiltext{4}{
  Department of Astronomy, University of Michigan, 500 Church St., Ann
  Arbor, MI 48109 USA
\label{Michigan}}

\altaffiltext{5}{
  Astrophysics Group, Imperial College London, Blackett Laboratory,
  Prince Consort Road, London SW7 2AZ UK
\label{Imperial}}

\altaffiltext{6}{
  Center for Cosmology and Particle Physics, Department of Physics,
  New York University, New York, NY 10003
\label{NYU}}

\altaffiltext{7}{
  Spitzer Science Center, Caltech, Pasadena, CA 91125 USA
\label{Spitzer}}

\altaffiltext{8}{
  National Optical Astronomical Observatories, Tucson, AZ 85719 USA
\label{NOAO}}

\altaffiltext{9}{
  Space Telescope Science Institute, 3700 San Martin Drive, Baltimore,
  MD 21218 USA
\label{STScI}}

\altaffiltext{10}{
  Institute for Astronomy, University of Edinburgh, Royal Observatory,
  Edinburgh, EH9 3HJ UK
\label{Edinburgh}}

\altaffiltext{11}{
  Department of Physics and Astronomy, Texas A\&M University, 4242
  TAMU, College Station, TX 77843 USA
\label{TexasAM}}

\altaffiltext{12}{
  Carnegie Observatories, 813 Santa Barbara Street, Pasadena, CA 91101
  USA
\label{Carnegie}}

\altaffiltext{13}{
  Max-Planck-Institut für extraterrestrische Physik,
  Giessenbachstrasse 1, D-85748 Garching bei München, Germany
\label{Max-Planck}}

\altaffiltext{14}{
  Department of Physics and Astronomy, University of Pittsburgh, 3941
  O'Hara St, Pittsburgh, PA 15260 USA
\label{Pitt}}

\altaffiltext{15}{
  Department of Physics and Astronomy, Johns Hopkins University,
  Baltimore, MD 21218 USA
\label{Johns Hopkins}}

\altaffiltext{16}{
  Astrophysics Science Division, Goddard Space Flight Center, Code
  665, Greenbelt, MD 20771 USA
\label{Goddard}}

\def\etal{et al.}
\newcommand{\Ha}{\hbox{{\rm H}$\alpha$}}
\newcommand{\Hb}{\hbox{{\rm H}$\beta$}}
\newcommand{\OII}{\hbox{[{\rm O}\kern 0.1em{\sc ii}]}}
\newcommand{\OIII}{\hbox{[{\rm O}\kern 0.1em{\sc iii}]}}
\newcommand{\NII}{\hbox{[{\rm N}\kern 0.1em{\sc ii}]}}
\newcommand{\SII}{\hbox{[{\rm S}\kern 0.1em{\sc ii}]}}

\begin{abstract}

We present {\it Hubble Space Telescope} Wide Field Camera 3 slitless
grism spectroscopy of 28 emission-line galaxies at $z \sim 2$, in the
GOODS-S region of the Cosmic Assembly Near-infrared Deep Extragalactic
Legacy Survey (CANDELS).  The high sensitivity of these grism
observations, with $>1\sigma$ detections of emission lines to $f>2.5
\times 10^{-18}$~erg~s$^{-1}$~cm$^{-2}$, means that the galaxies in
the sample are typically $\sim$7 times less massive (median
$M_*=10^{9.5}M_{\odot}$) than previously studied $z \sim 2$
emission-line galaxies.  Despite their lower mass, the galaxies have
\OIII/\Hb\ ratios which are very similar to previously studied $z \sim
2$ galaxies and much higher than the typical emission-line ratios of
local galaxies.  The WFC3 grism allows for unique studies of spatial
gradients in emission lines, and we stack the two-dimensional spectra
of the galaxies for this purpose.  In the stacked data the \OIII\
emission line is more spatially concentrated than the \Hb\ emission
line with 98.1\% confidence.  We additionally stack the X-ray data
(all sources are individually undetected), and find that the average
$L_{\rm [OIII]}/L_{\rm 0.5-10 keV}$ ratio is intermediate between
typical $z \sim 0$ obscured active galaxies and star-forming galaxies.
Together the compactness of the stacked \OIII\ spatial profile and the
stacked X-ray data suggest that at least some of these low-mass,
low-metallicity galaxies harbor weak active galactic nuclei.

\end{abstract}


\section{Introduction}

Rest-frame optical spectra contain a wealth of information about
galaxy properties.  In particular, the strengths and ratios between
collisionally excited and recombination emission lines can be used to
infer star formation rate \citep[SFR, e.g.][]{ken98}, gas-phase
metallicity \citep[e.g.][]{kew01}, and nuclear activity
\citep[e.g.][]{bpt81,kew06}.  However applying these techniques to
galaxies at higher redshifts is challenging for two reasons: (1) at
$z>1.5$ optical emission lines are observed in the near-infrared
(NIR), where ground-based spectroscopy is challenging, and (2) the
characteristics of $z>1$ galaxies may be physically different from
local galaxies.

\begin{figure}
\scalebox{1.2}
{\plotone{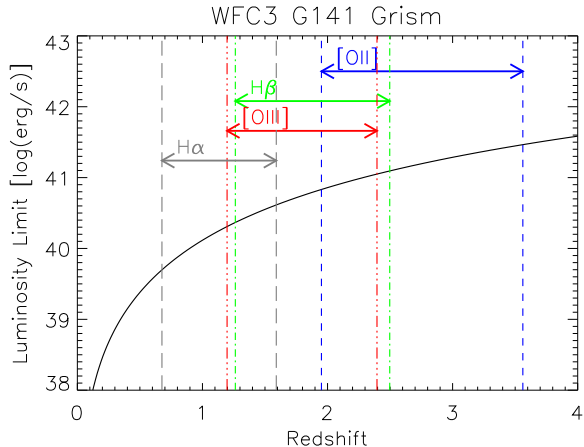}}
\figcaption{Line luminosity limits for detecting the \Ha, \Hb, \OIII,
  and \OII\ emission lines with 8 orbits of the HST/WFC3 G141 grism
  (corresponding to a line flux limit of $2.5 \times
  10^{-18}$~erg/s/cm$^2$).  Dashed vertical lines indicate the
  redshift ranges at which each line is observed in the G141
  wavelength range ($1.1<\lambda<1.7\mu$m).  In this paper we study
  galaxies with \Hb\ and \OIII\ lines, selected at $1.3<z<2.4$.
\label{fig:grismlines}}
\end{figure}

The Wide Field Camera 3 (WFC3) G141 slitless grism on the {\it Hubble
Space Telescope} ({\it HST}) addresses the first problem, with far
improved capabilities for NIR spectroscopy than ground-based
observatories \citep[e.g.,][]{str11}.  Most of the previous
ground-based studies of $z>1.5$ galaxies have been biased to massive
and UV-luminous systems \citep{shap04,erb06,hay09,ono10}, probably
missing most of the more numerous, less massive galaxies at $1.5<z<3$.
Only a handful of lower-mass galaxies at $z \sim 2$ have been studied,
revealing much lower metallicities than both $z \sim 0$ galaxies of
the same mass and their massive and evolved counterparts at the same
redshift \citep{yuan09,erb10,fin11}.  In contrast the tremendous
sensitivity of {\it HST}/WFC3 allows for $>1\sigma$ detections of
emission lines to $f>2.5 \times 10^{-18}$~erg~s$^{-1}$~cm$^{-2}$ for
every object in a $2\farcm1 \times 2\farcm1$ field with 8 orbits.  In
addition, {\it HST} is not limited by atmospheric emission or
absorption in the NIR, and so spectral lines are detectable over the
entire 1.1-1.7 $\mu$m wavelength range of the WFC3 G141 grism.  Figure
\ref{fig:grismlines} shows the detectability of emission lines with
the G141 grism at various redshift ranges.

Star formation processes were quite different at $z>1$ than in the
local universe \citep[e.g.][]{pap05,red06,shim11}, and so it is not
clear if the emission-line diagnostics used to characterize local
galaxies can be directly applied to $z>1$ systems.  For example, the
bulk of $z \sim 2$ galaxies have higher ratios of collisionally
excited to recombination lines ($\OIII/\Hb$ and $\NII/\Ha$) than local
galaxies \citep{shap05b,erb06,kri07}.  Local galaxies with high SFR
tend to have higher emission-line ratios than galaxies with lower SFR,
suggesting that rapid star formation leads to different ionization
properties \citep{liu08,bri08}.  Since $z>1$ galaxies typically have
higher SFR than local analogs of the same mass \citep[e.g.][]{pap05},
their higher line ratios may simply be the result of their higher SFR.
However \citet{wri10} used spatially resolved spectroscopy of one
galaxy at $z=1.6$ to show that these line ratios increase in the
center of the galaxy, with normal star-forming (SF) line ratios at
outer apertures.  This suggests that weak or obscured active galactic
nucleus (AGN) activity may be the cause of higher line ratios in $z
\sim 2$ galaxies, perhaps because AGN activity is correlated with SFR
as observed locally \citep{kau03,shi09}.  There are many predictions
for the density of obscured and Compton-thick (CT, $N_H \gtrsim
10^{24}$~cm$^2$) AGNs at $1.5<z<3$ from the X-ray background
\citep{gil07,tre09} and from IR-excess galaxy counts and X-ray
stacking \citep{dad07,mar07,fio09,geo11}, but to date only three CT
AGNs have been individually confirmed at $z \sim 2$
\citep{ale05,erl08,ale08}.  Obscured AGNs are missed by even the
deepest X-ray surveys, but can still exhibit the narrow emission-line
signature of an AGN \citep[e.g.,][]{lam11}.  Measuring the density of
obscured AGNs at $z \sim 2$ is especially important to reveal how much
black hole growth occurs in an enshrouded phase.

Here we use 8.5 orbits of WFC3 G141 spectroscopy to study the
properties of emission-line galaxies at $1.3<z<2.4$ in the Hubble
Ultra Deep Field \citep[HUDF,][]{bec06}.  These data were taken as
part of the supernova follow-up program in the Cosmic Assembly
Near-Infrared Deep Extragalactic Legacy Survey
\citep[CANDELS\footnote{http://candels.ucolick.org},][]{gro11,koe11}.
The supernova candidate identified in the HUDF region is described by
\citet{rod11}, and this work makes use of the spectra that were
serendipitously obtained for hundreds of other targets in the
$2.1\arcmin \times 2.1\arcmin$ field.  In particular we focus on the
28 galaxies exhibiting \Hb\ and \OIII\ emission lines in the observed
grism wavelength range (at $1.3<z<2.4$).  Like the study by
\citet{wri10}, the high spatial resolution of the {\it HST}/WFC3
similarly allows for studies of spatial gradients in emission-line
properties.  Unlike an integral field unit spectrograph, however, the
WFC3 grism has the unique ability to do this for hundreds of objects
in the field of view simultaneously.

In \S 2 we describe the observations, data reduction, and emission
line luminosity limits.  In \S 3 we describe our method for
calculating emission lines, complicated by the limited spectral
resolution of the WFC3 G141 grism.  In \S 4 we stack the spatially
resolved spectra to show that both the emission-line properties and
the stacked X-ray data are suggestive of weak AGN activity in at least
some of these galaxies.  We summarize our results in \S 5.  Throughout
the paper we adopt a cosmology with $h=0.70$, $\Omega_M=0.3$,
$\Omega_{\Lambda}=0.7$.

\section{HST/WFC3 Slitless Grism Observations}

The HUDF was observed with the {\it HST}/WFC3 G141 grism on 26-27 Oct
2010 and 1 Nov 2010 for a total of 8.5 orbits.  The data were acquired
in two different orients, with 2.5 orbits at an orient of
$PA=176^\circ$ on the sky and 6 orbits at an orient of
$PA=182.4^\circ$.  The data for each orient were reduced using the aXe
software \citep[][available at {\tt
http://axe.stsci.edu/axe/}]{kum09}, which flux- and
wavelength-calibrated the data and provided co-added two-dimensional
(2D) spectra as well as optimally extracted one-dimensional (1D)
spectra.  The usable wavelength range of each reduced spectrum is
$1.1<\lambda<1.7\mu$m, with a resolution of $R \simeq 130$
(46.5\AA/pixel for a point source and 0\farcs13/pixel).

The G141 grism is slitless, and this introduces a number of
complexities when interpreting the data.  Spatially extended sources
have features which appear extended in the wavelength direction, and
so broad emission or absorption lines can be caused by either large
velocity widths or by the size of the object.  In addition, spectra
frequently suffer from contamination by 0th order, 1st order, and 2nd
order spectra of nearby sources.  For this reason we did not combine
the two orients, instead analyzing each orient independently and using
them to identify contamination.

We calculated redshifts for all 455 targets detected in the HUDF G141
data.  The best redshift was found using a cross-correlation
template-fitting IDL algorithm based on the publicly available {\tt
idlspec2d} package written by
D. Schlegel\footnote{http://spectro.princeton.edu/idlspec2d\_install.html}.
Previous spectroscopic redshifts \citep{wolf04,mig05,van08,str08} and
photometric redshifts \citep{wolf04,gra06,ryan07,wuy08,car10} were
used as priors.  We used two templates: a star-forming emission-line
galaxy and a passive absorption line galaxy.  Each template was
constructed using the high-resolution optical ($3800<\lambda<9200$\AA)
templates from the SDSS
library\footnote{http://www.sdss.org/dr5/algorithms/spectemplates/}
combined with supplemental low-resolution UV and NIR data from the
SWIRE photometric templates \citep{pol07}, with the final templates
resampled to match the spectral resolution of the WFC3 G141 data.  All
spectra and their redshift fits were visually inspected to ensure
quality.

\begin{deluxetable*}{rcccccccccc}
%\rotate
%\tabletypesize{\scriptsize}
\tablecolumns{11}
\tablecaption{HUDF WFC3/G141 Emission-Line Galaxies\label{tbl:targets}}
\tablehead{
  \colhead{\#} & 
  \colhead{RA} & 
  \colhead{Dec} & 
  \colhead{$z$} & 
  \colhead{$H_{\rm AB}$} & 
  \colhead{$\log(L_{\Hb})$} & 
  \colhead{$\log(L_{\OIII})$} & 
  \colhead{$f_{\OIII}/f_{\Hb}$} & 
  \colhead{$(U-B)_{\rm rest}$} &
  \colhead{$\log(M_*)$} &
  \colhead{$f_{\rm 0.5-10 keV}$} \\
  \colhead{-} & 
  \colhead{[deg]} & 
  \colhead{(J2000)} & 
  \colhead{-} & 
  \colhead{[mag]} & 
  \colhead{[log(erg/s)]} & 
  \colhead{[log(erg/s)]} & 
  \colhead{-} & 
  \colhead{[mag]} &
  \colhead{$[\log(M_{\odot})]$} &
  \colhead{[$10^{-16}$~erg~s$^{-1}$~cm$^{-2}$]} }
\startdata
 1 & 53.135250 & -27.781778 &  1.44 & 22.67 & $41.08_{-0.15}^{+0.28}$ & $41.48_{-0.12}^{+0.14}$ & $ 2.52_{-1.13}^{+0.77}$ &  0.76 & 10.39 & $<0.45$ \\
 2 & 53.141159 & -27.759295 &  1.77 & 24.19 & $40.89_{-0.11}^{+0.21}$ & $41.61_{-0.05}^{+0.05}$ & $ 5.23_{-2.06}^{+1.22}$ &  0.51 &  9.39 & $<0.44$ \\
 3 & 53.143940 & -27.779781 &  1.71 & 25.21 & $40.92_{-0.16}^{+0.25}$ & $41.35_{-0.21}^{+0.07}$ & $ 2.97_{-1.77}^{+0.18}$ &  0.38 &  9.05 & $<0.47$ \\
 4 & 53.144127 & -27.773602 &  1.90 & 23.58 & $41.18_{-0.18}^{+0.11}$ & $41.92_{-0.02}^{+0.05}$ & $ 5.54_{-1.14}^{+2.72}$ &  0.36 &  9.51 & $<0.44$ \\
 5 & 53.144485 & -27.791162 &  1.42 & 23.52 & $40.97_{-0.02}^{+0.28}$ & $41.93_{-0.05}^{+0.05}$ & $ 8.09_{-3.35}^{+1.05}$ &  0.46 &  9.20 & $<0.46$ \\
 6 & 53.144840 & -27.778469 &  2.22 & 24.37 & $41.10_{-0.12}^{+0.23}$ & $41.95_{-0.06}^{+0.02}$ & $ 7.10_{-3.24}^{+1.45}$ &  1.00 &  9.28 & $<0.47$ \\
 7 & 53.146435 & -27.778353 &  1.87 & 24.74 & $41.18_{-0.22}^{+0.10}$ & $41.94_{-0.09}^{+0.01}$ & $ 5.74_{-1.61}^{+2.62}$ & -0.01 &  8.43 & $<0.47$ \\
 8 & 53.146900 & -27.781837 &  1.55 & 24.09 & $40.74_{-0.14}^{+0.23}$ & $41.27_{-0.13}^{+0.10}$ & $ 3.35_{-1.51}^{+0.97}$ &  0.48 &  9.39 & $<0.46$ \\
 9 & 53.147469 & -27.777693 &  1.83 & 22.80 & $40.83_{-0.05}^{+0.33}$ & $41.57_{-0.10}^{+0.06}$ & $ 4.52_{-2.15}^{+1.11}$ &  1.05 & 10.43 & $<0.47$ \\
10 & 53.147747 & -27.765675 &  2.31 & 23.75 & $41.08_{-0.08}^{+0.28}$ & $41.70_{-0.01}^{+0.21}$ & $ 4.22_{-1.61}^{+1.96}$ &  0.59 &  9.91 & $<0.43$ \\
11 & 53.147804 & -27.771420 &  2.17 & 24.52 & $41.36_{-0.28}^{+0.07}$ & $42.01_{-0.09}^{+0.04}$ & $ 5.09_{-1.24}^{+2.35}$ &  0.65 &  8.82 & $<0.45$ \\
12 & 53.148003 & -27.787737 &  1.90 & 23.80 & $40.96_{-0.11}^{+0.31}$ & $41.24_{-0.16}^{+0.19}$ & $ 2.02_{-1.09}^{+0.35}$ &  0.63 &  9.92 & $<0.45$ \\
13 & 53.149063 & -27.785116 &  2.07 & 24.20 & $41.02_{-0.23}^{+0.25}$ & $41.38_{-0.23}^{+0.10}$ & $ 1.90_{-0.88}^{+1.28}$ &  0.65 &  9.72 & $<0.44$ \\
14 & 53.151283 & -27.792444 &  1.86 & 24.05 & $41.18_{-0.05}^{+0.30}$ & $41.74_{-0.04}^{+0.15}$ & $ 3.62_{-1.62}^{+0.77}$ &  0.48 &  9.59 & $<0.43$ \\
15 & 53.151691 & -27.763506 &  2.16 & 25.19 & $40.55_{-0.04}^{+0.39}$ & $41.24_{-0.33}^{+0.03}$ & $ 2.88_{-1.80}^{+1.06}$ &  0.48 &  8.94 & $<0.44$ \\
16 & 53.152264 & -27.770132 &  1.85 & 23.21 & $41.26_{-0.24}^{+0.13}$ & $41.77_{-0.04}^{+0.06}$ & $ 3.23_{-0.88}^{+2.34}$ &  0.53 &  9.97 & $<0.47$ \\
17 & 53.152870 & -27.772545 &  1.85 & 24.13 & $41.29_{-0.21}^{+0.03}$ & $41.95_{-0.04}^{+0.03}$ & $ 5.52_{-1.36}^{+1.58}$ &  0.36 &  9.16 & $<0.46$ \\
18 & 53.152874 & -27.780163 &  1.86 & 24.33 & $40.65_{-0.12}^{+0.26}$ & $41.32_{-0.21}^{+0.01}$ & $ 3.73_{-1.72}^{+0.90}$ &  0.60 &  9.46 & $<0.46$ \\
19 & 53.153774 & -27.767345 &  2.32 & 23.65 & $40.98_{-0.16}^{+0.22}$ & $41.58_{-0.11}^{+0.11}$ & $ 3.94_{-1.61}^{+1.50}$ &  0.44 &  9.77 & $<0.46$ \\
20 & 53.154453 & -27.771494 &  2.23 & 22.76 & $41.67_{-0.09}^{+0.14}$ & $42.49_{-0.06}^{+0.01}$ & $ 6.63_{-2.31}^{+1.01}$ &  0.66 & 10.24 & $<0.46$ \\
21 & 53.155643 & -27.779324 &  1.85 & 22.10 & $41.60_{-0.18}^{+0.22}$ & $41.83_{-0.12}^{+0.13}$ & $ 1.27_{-0.32}^{+1.33}$ &  0.76 & 10.72 & $<0.45$ \\
22 & 53.156372 & -27.767874 &  2.02 & 24.07 & $40.75_{-0.05}^{+0.34}$ & $41.23_{-0.20}^{+0.12}$ & $ 2.61_{-1.39}^{+0.31}$ &  0.83 &  9.19 & $<0.47$ \\
23 & 53.157207 & -27.778561 &  1.31 & 23.86 & $40.94_{-0.15}^{+0.31}$ & $41.45_{-0.28}^{+0.08}$ & $ 3.17_{-1.88}^{+0.01}$ &  0.54 &  9.20 & $<0.45$ \\
24 & 53.160435 & -27.775246 &  1.41 & 20.86 & $41.08_{-0.26}^{+0.19}$ & $41.34_{-0.04}^{+0.07}$ & $ 1.81_{-0.62}^{+1.34}$ &  0.42 &  9.64 & $<0.46$ \\
25 & 53.160446 & -27.790394 &  1.61 & 23.44 & $41.23_{-0.50}^{+0.15}$ & $41.57_{-0.12}^{+0.07}$ & $ 1.39_{-0.11}^{+4.25}$ &  0.48 &  9.74 & $<0.43$ \\
26 & 53.165157 & -27.781664 &  2.22 & 24.56 & $40.90_{-0.15}^{+0.25}$ & $41.60_{-0.19}^{+0.07}$ & $ 5.88_{-3.79}^{+0.28}$ &  0.69 &  9.32 & $<0.44$ \\
27 & 53.172153 & -27.763655 &  2.21 & 24.83 & $41.03_{-0.23}^{+0.23}$ & $41.79_{-0.11}^{+0.14}$ & $ 5.77_{-1.84}^{+3.09}$ &  0.75 &  9.04 & $<0.48$ \\
28 & 53.172241 & -27.760641 &  1.54 & 21.83 & $41.33_{-0.07}^{+0.56}$ & $41.65_{-0.06}^{+0.19}$ & $ 2.09_{-1.44}^{+0.40}$ &  0.72 & 10.75 & $<0.48$ \\
\enddata
\end{deluxetable*}

Of the 455 targets, we selected the 28 galaxies at $1.3<z<2.4$ and
\Hb\ or \OIII\ emission-line luminosities of $L>1.8 \times
10^{41}$~erg~s$^{-1}$.  This luminosity limit represents the flux
limit for detecting emission lines at the $1\sigma$ level ($f>2.5
\times 10^{-18}$~erg~s$^{-1}$) at $z=2.4$.  For all 28 galaxies we
detect both \Hb\ and \OIII\ in both grism pointings, and as a result
their redshifts are unambiguous.  Table \ref{tbl:targets} shows the
properties of these 28 emission-line galaxies.  Their median redshift
is 1.86.

\subsection{Stellar Masses and Rest-frame Colors}

Rest-frame magnitudes and stellar masses were calculated from the
GOODS-MUSIC photometry catalog \citep{san09} by fitting a set of
template spectral energy distributions drawn from the PEGASE stellar
population models (see \citealp{pegase} for a description of an
earlier version of this stellar population model).  We allow dust
attenuation following \citet{cal01}, with values of $E(B-V)_{\rm gas}$
between -0.05 and 1.5.  (Slightly negative $E(B-V)$ avoids biasing the
mean properties of the population by accounting for the rare cases
when photometric error gives fits that prefer slightly negative
attenuation.  In practice all of our emission-line galaxies have
best-fit $E(B-V)>0$ anyway.)  All templates have solar metallicity,
and included a broad range of exponentially-decreasing, constant, or
exponentially-rising star formation beginning at $z_f \sim 4$.  A grid
of stellar population templates were compared with photometric data
points of each galaxy, given the spectroscopic redshift from the grism
data, and the best matching template was used to calculate rest-frame
colors and a stellar $M/L$ ratio.  Compared to galaxies with
independently-estimated stellar masses (from independent photometry of
similar but not identical datasets) in the GOODS-S field
\citep{wuy08}, our method yields similar masses for intensely
star-forming galaxies both with and without dust attenuation (i.e.,
the galaxies of interest here), with a scatter of 0.3 dex.  Our masses
are 0.1-0.25 dex larger for moderately star-forming galaxies or
non-star-forming galaxies because our templates include older stellar
populations than those used by \citet{wuy08}.  Rest-frame colors are
reproduced to within 0.1 mag in rest-frame $(U-B)$ color for all
galaxy types.  Full details on the derived masses and rest-frame
quantities are given by \citet{bell11}.

As shown by \citet{vanderwel11}, strong emission lines can have
significant effects on broad-band colors in $z \sim 2$ galaxies.  In
our galaxies the median total equivalent width (EW) of the \Hb\ and
\OIII\ lines is $EW_{\rm obs} = 220$\AA: this brightens the rest-frame
$B$ magnitude by $\sim20\%$ and reddens the $(U-B)$ color by $0.2$
magnitudes.  Despite these effects, we do not correct rest-frame
magnitudes for emission lines for two reasons.  First, the \OII\
emission line could have the opposite effect, brightening $U$ and
causing $U-B$ to appear bluer, but we do not measure this emission
line because it is not generally present in our observed spectral
range.  In addition, the SF/AGN diagnostic we use \citep{yan11} was
calibrated with $(U-B)$ colors uncorrected for emission lines, and so
we do not correct our colors to remain consistent.

\begin{figure}
\epsscale{1.2}
{\plotone{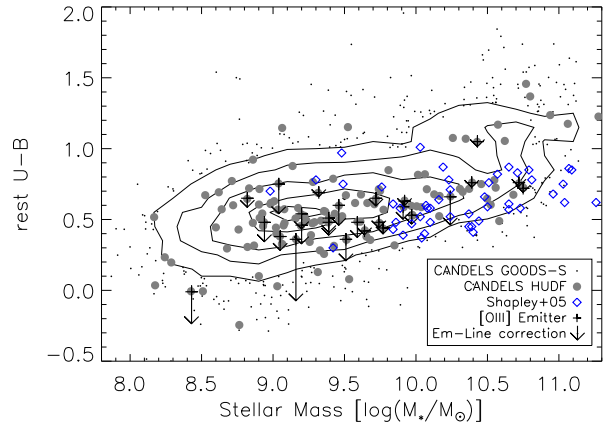}}
\figcaption{Rest-frame $(U-B)$ color and stellar mass for our 28
  emission-line galaxies (black crosses).  Typical $1\sigma$ errors
  are $\sim$0.1 magnitudes in $(U-B)$ and $\sim$0.3~dex in mass.  For
  comparison filled gray circles show the 137 galaxies at $1.3<z<2.4$
  in the HUDF and gray contours and points show the entire set of 4140
  GOODS-S galaxies in the redshift range, both with masses and
  rest-frame quantities calculated from the same photometry and with
  the same method.  Blue diamonds show a comparison sample of $z \sim
  2$ galaxies from \citet{shap05a}: this is the parent sample of most
  previous studies of emission-line galaxies at $z \sim 2$.  Arrows
  show the effects on $(U-B)$ color if the \Hb\ and \OIII\ emission
  lines are removed (although the \OII\ emission line, which is not
  measured here, could have the opposite effect on $(U-B)$ color).
  The WFC3 grism reveals galaxies with masses a factor of $\sim7$
  lower than previous studies of $z \sim 2$ emission-line galaxies.
\label{fig:colormass}}
\end{figure}

Figure \ref{fig:colormass} shows the rest-frame $(U-B)$ colors and
stellar masses for our 28 emission-line galaxies.  Also shown by gray
filled circles are the 137 galaxies in the HUDF with photometric or
spectroscopic redshifts at the same $1.3<z<2.4$, with rest-frame
colors and stellar masses computed in an identical fashion.  The
emission-line galaxies studied here are rather common at $1.3<z<2.4$,
making up $\sim$20\% of the total $H<26$ galaxy population.  Gray
contours additionally show the set of all 4140 GOODS-S galaxies with
photometric or spectroscopic redshifts in the same $1.3<z<2.4$
redshift range, and our galaxies have a roughly similar distribution
of mass and slightly bluer colors than the larger parent distribution.
For comparison with previous samples, we also show the 50 galaxies at
$1.3<z<2.4$ of \citet{shap05a}, using their masses and $(U-B)$ colors
calculated from their photometry using the publicly available {\tt
kcorrect} package \citep{kcorrect}.  The galaxies of \citet{shap05a}
form the parent sample for most previous $z \sim 2$ emission-line
studies \citep[e.g.,][]{shap05b,erb06,erb10,wri10}.  The median mass
of the grism galaxies is $M_* \sim 10^{9.5}M_{\odot}$, a factor of
$\sim7$ lower than the median mass of the \citet{shap05a} sample.
Several of the emission-line galaxies from the WFC3 grism data appear
to be star-forming dwarf galaxies with masses of only $M_* \sim
10^{9}M_{\odot}$.

\subsection{X-ray Data}

The HUDF has the deepest X-ray data in the sky, with 4 Ms of {\it
Chandra} data.  Imaging was obtained in 54 observations (obsIDs) over
the course of 3 {\it Chandra} observing cycles in 2000, 2007 and 2010
using the Advanced CCD Imaging Spectrometer imaging array
\citep[ACIS-I][]{gar03}.  The data were reduced using CIAO v4.2
according to the basic procedure described in \citet{lai09}.  Before
combining the observations, the astrometry of each obsID was
registered to that of the GOODS-MUSYC survey \citep{gaw06} by matching
the positions of bright X-Ray sources to H-band selected sources,
using the tool {\tt reproject\_aspect}.  Source detection was carried
out according to the method described in \citet{lai09}, and none of
our galaxies are detected.  We instead assign $3\sigma$ upper limits
calculated using the background and PSF size from the location of each
galaxy on the {\it Chandra}/ACIS detector.

%% One of our 34 $z \sim 2$ emission line galaxies was detected in the
%% 4 Ms X-ray data, with a flux of $f_{\rm 0.5-10 keV} = 9.6 \times
%% 10^{-17}$~erg~s$^{-1}$~cm$^{-2}$ ($L_{\rm 0.5-10 keV} = 2.7 \times
%% 10^{42}$~erg~s$^{-1}$).

\section{Line Measurements}

Because the grism is slitless, any spatial extent along the dispersion
direction will artificially broaden spectral features.  The
\OIII$\lambda4959$ and \OIII$\lambda5007$ lines are often completely
blended together in resolved galaxies, and larger systems additionally
have \Hb$\lambda4861$ partially blended with the two \OIII\ lines.
For this reason we do not attempt to individually measure each of the
\Hb\ \OIII\ lines directly from the data.  Instead we subtract a
linear continuum in the region $4770<\lambda<5100$\AA\ and fit the
residual with a mixture of three Gaussians.  The three Gaussians are
constrained to be centered within 20\AA\ ($\sim$1200 km~s$^{-1}$) of
4861\AA, 4959\AA, and 5007\AA, in the rest frame.  Because the two
\OIII\ lines are blended, we measure the \OIII$\lambda$5007 flux as
3/4 of the total \OIII\ flux (i.e., 3/4 of the the sum of the two
Gaussians centered near 4959\AA\ and 5007\AA; \citealt{sto00}).
Errors in line flux (and $f(\OIII)/f(\Hb)$ and $L_{\rm [OIII]}$) are
calculated by bootstrapping 10~000 realizations of the resampled data.

\begin{figure*} %figure*
\plotone{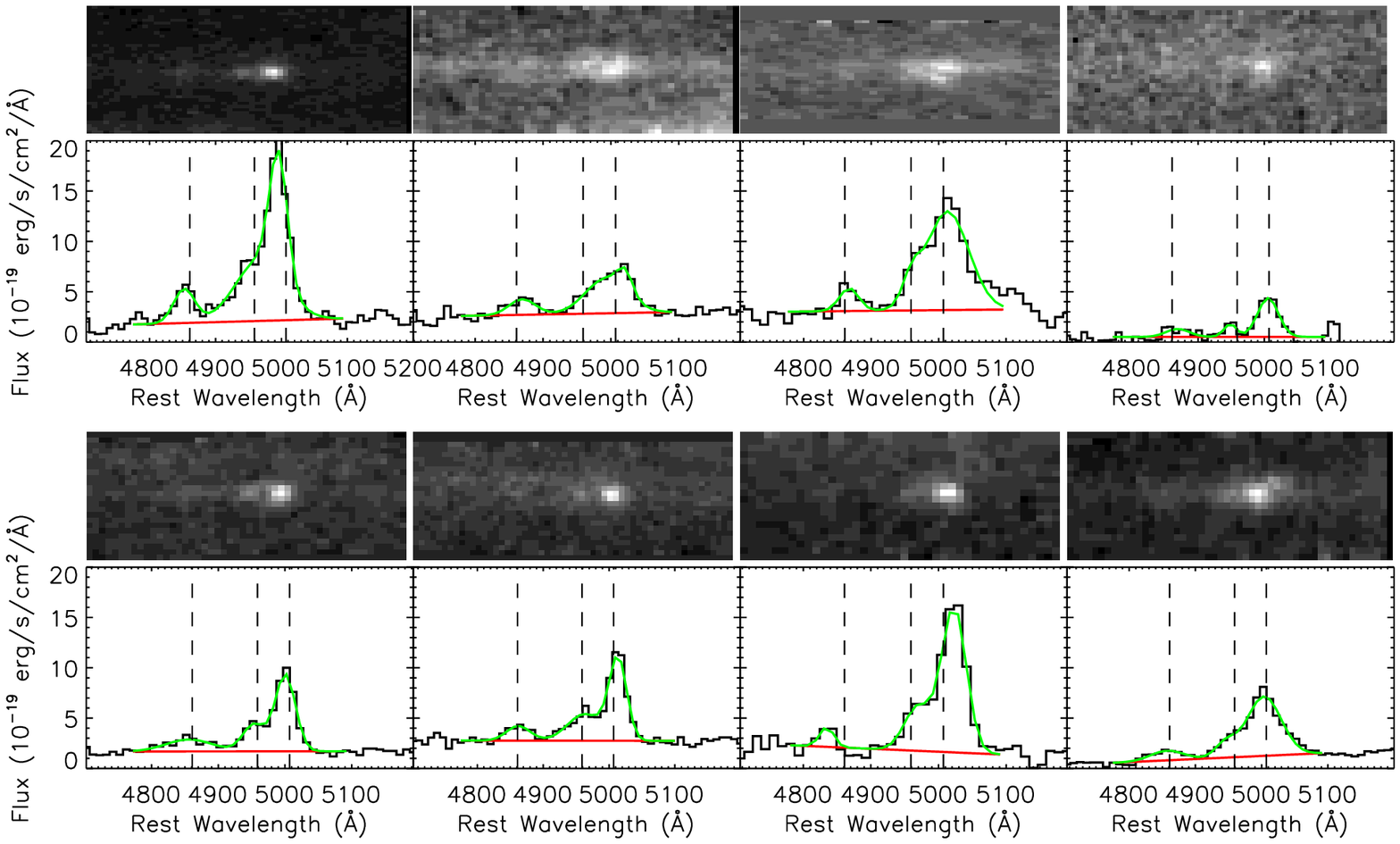}
\figcaption{Examples of fits to the \Hb\ and \OIII\ emission lines for
  eight of the emission-line galaxies.  Dashed vertical lines indicate
  the line centers for the \Hb$\lambda4861$, \OIII$\lambda4959$, and
  \OIII$\lambda5007$ emission lines.  Because the G141 grism has such
  low spectral resolution, the \Hb\ and \OIII\ lines are frequently
  blended.  We subtract the continuum, shown by the red line, and fit
  the entire system with a mixture of three Gaussians, shown by the
  green line.  The \OIII$\lambda5007$ flux is measured as 3/4 of the
  total \OIII\ flux.
\label{fig:fitplot}}
\end{figure*}

Figure \ref{fig:fitplot} shows the \Hb\ and \OIII\ line regions for
eight emission-line galaxies.  The 2D dispersed image, 1D spectrum,
and line fit is shown for each object.

The classical ``BPT'' diagnostic \citep{bpt81} uses emission-line
ratios (namely, $f(\OIII)/f(\Hb)$ and $f(\NII)/f(\Ha)$) to distinguish
galaxies dominated by star formation or black hole activity in the
local universe \citep[see also][]{vei87,kew01}.  However in the WFC3
grism data, \Ha\ and \NII\ are hopelessly blended, and additionally
are redshifted beyond the observed grism redshift range at $z>1.6$.
\citet{yan11} showed that it is possible to construct an alternate
version of the BPT diagnostic which replaces $f(\NII)/f(\Ha)$ with
rest-frame $(U-B)$ color.  This modified BPT diagnostic is especially
useful for interpreting the WFC3 grism data.

An alternative SF/AGN classification diagnostic possible for our
galaxies is the ``MEx'' diagram \citep{jun11}, which replaces the
$(U-B)$ color with stellar mass $M_*$.  Stellar mass, however, is
dependent on the physics of the model, and could be systematically
inaccurate due to the constraints of available templates.  Meanwhile
rest-frame color does not depend on the physics of the model, only its
goodness of fit to the data.  For this reason we prefer $(U-B)$ over
$M_*$, noting that the \citet{yan11} diagnostic agrees well with the
``MEx'' method \citep[see Appendix A of][]{jun11}.

\begin{figure*} %figure*
\plotone{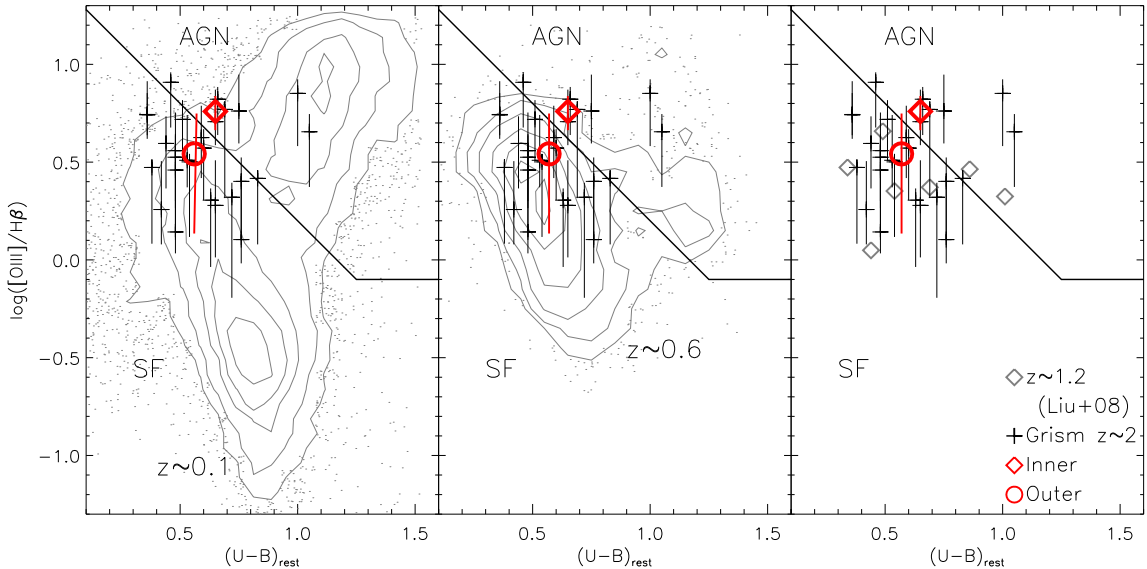}
\figcaption{Ratio of \OIII\ to \Hb\ with rest-frame $(U-B)$ color.
  The solid line shows the division between AGN (upper right) and star
  formation (left and bottom) from \citet{yan11}.  Crosses show the 28
  $z \sim 2$ galaxies from the WFC3 G141 data.
%% , with the X-ray detected source indicated by a blue circle.
  Contours and points show comparison samples \citep[also
  from][]{yan11} at $z \sim 0$ (left panel) and $z \sim 0.6$ (middle
  panel).  In the right panel, gray diamonds show a comparison sample
  at $z \sim 1.2$ from \citet{liu08}.  The red points show the colors
  and line ratios of the stacked data: the diamond shows the inner
  $0\farcs39$ region and the circle shows the outer
  $0\farcs39-0\farcs78$ region.  The $z \sim 0$ and $z \sim 0.6$
  comparison samples have the same line luminosity constraint as the
  WFC3 G141 data, with either \Hb\ or \OIII\ of $L>1.8 \times
  10^{41}$~erg~s$^{-1}$.  From $z \sim 0.1$ to $z \sim 0.6$ to $z \sim
  1.2$, galaxies have lower metallicities and consequently move to the
  upper left of the figure, and our $z \sim 2$ galaxies lie in very
  similar space to the previously studied $z \sim 1.2$ galaxies of
  \citet{liu08}.  The line ratio and color of the inner region of the
  stacked data is marginally ($1.0\sigma$) more AGN-like than the
  outer region.
\label{fig:bpt}}
\end{figure*}

Figure \ref{fig:bpt} shows our $z \sim 2$ emission-line galaxies on
the modified BPT diagnostic of \citet{yan11}, along with comparison
samples at $z \sim 0$ and $z \sim 0.6$ \citep[also from][]{yan11} and
$z \sim 1.2$ \citep[from][]{liu08}.  Galaxies dwelling on the left and
bottom of each panel tend to be SF dominated, while systems in the
upper right are AGN dominated.  Low metallicity galaxies tend to have
bluer colors and higher $f(\OIII)/f(\Hb)$ ratios, and indeed the locus
of SF galaxies tends to shift to the upper left from $z \sim 0$ to $z
\sim 0.6$ to $z>1$.  Our galaxies generally lie in same region as
previously studied $z>1$ galaxies.

Directly interpreting the position of $z>1$ galaxies on the modified
BPT diagnostic in Figure \ref{fig:bpt} is challenging for several
reasons, however.  First, the dividing line of \citet{yan11} between
SF galaxies and AGN was empirically calibrated at $z<1$ only, and it
is unclear if this dividing line changes with redshift.  Local
galaxies with higher SFR tend to have higher line ratios
\citep{liu08,bri08}, and since $z \sim 2$ galaxies have higher SFR
galaxies than local galaxies of similar mass \citep{pap05}, we might
already expect them to have a higher $f(\OIII)/f(\Hb)$ ratios from
both their rapid star formation and low metallicity.  On the other
hand, active galaxies might also be more likely to be confused as
SF-dominated in low-mass, low-metallicity galaxies.  \citet{izo08}
showed that AGN accreting as high as the Eddington limit contribute
less than 10\% to the emission-line ratios in metal-poor dwarf
galaxies.  Our $z \sim 2$ galaxies have similarly low masses, and the
emission from their low-mass black holes would be overwhelmed by a SFR
of a few solar masses per year.

For these reason it is difficult to interpret the absolute position of
objects, and more instructive to consider relative positions of the
inner and outer regions of the galaxies, as well as differences in the
profiles of \OIII\ and \Hb.

\subsection{Stacked {\it HST}/WFC3 Data}

The slitless WFC3 grism provides spatially resolved spectroscopy, with
a scale of $0\farcs13$ per pixel.  This is particularly useful for
studying gradients of emission-line ratios: most local galaxies have a
higher ratio of $f(\OIII)/f(\Hb)$ in outer regions due to negative
metallicity gradients \citep[e.g.,][]{vanzee98}, while a higher ratio
in a galaxy nucleus would suggest the presence of an AGN.  In general
the signal-to-noise (S/N) of our data is too low to measure line ratio
gradients for individual galaxies; instead we stack the 2D
spectroscopy to study line ratio gradients.

We stack the 2D images for each galaxy, equal-weighting each galaxy by
normalizing its total ($\Hb+\OIII\lambda4959+\OIII\lambda5007$) line
emission.  Images are aligned by the peak of the emission (dominated
by the \OIII$\lambda$5007 line).  This stacked spectrum allows us to
study the spatial extent of the \Hb\ and \OIII\ emission lines.
%% (We do not include the X-ray galaxy in the stack because its AGN is
%% already detected, while the stacked data might indicate the presence
%% of buried AGN in the X-ray-undetected sources.)  

\begin{figure}
\epsscale{1.2}
{\plotone{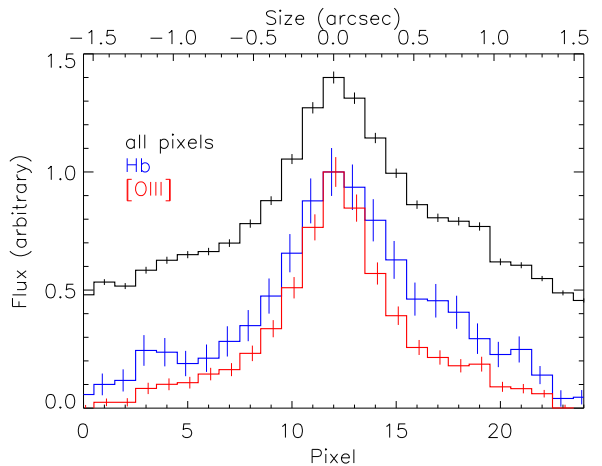}}
\figcaption{Spatial profiles of the \Hb\ (blue line) and \OIII\ (red
  line) emission lines, along with the entire spectrum (black line,
  offset by +0.4 in flux).  Each 1D profile is extracted along the
  cross-dispersion direction, using the three pixels (140\AA) about
  each line center for the emission lines or the entire wavelength
  range for the entire spectrum profile.  The \Hb\ profile is more
  extended than the \OIII\ profile with 98.1\% confidence, suggesting
  that at least some of the objects have a compact AGN contributing to
  the \OIII\ emission.
\label{fig:linespatial}}
\end{figure}

One-dimensional spatial profiles can be constructed by extracting each
emission line along the cross-dispersion direction.  Figure
\ref{fig:linespatial} shows the spatial profiles for the
\Hb$\lambda4861$ and \OIII$\lambda5007$ emission lines, along with the
collapsed cross-dispersion profile of the entire 2D spectrum.  Each
emission line is extracted from the central three pixels (140\AA)
about the line center.  The \Hb\ line has a cross-dispersion
full-width half-maximum (FWHM) of 7.2 pixels ($0\farcs94$,
$\sim$8~kpc), while the \OIII\ line is narrower with a FWHM of 4.9
pixels ($0\farcs64$, $\sim$5~kpc).

We test the significance of the \OIII\ line being more spatially
compact than the \Hb\ line by bootstrapping 100~000 resampled
profiles.  We construct each bootstrapped profile by adding a
normally-distributed random error (with $\sigma$ given by the
propagated error of the stacked profile) to the original data, and
then measure FWHM(\OIII) and FWHM(\Hb).  The fraction of bootstrapped
profiles with FWHM(\OIII)<FWHM(\Hb) gives the significance of \OIII\
being more compact than \Hb.  This occurs in 98.1\% of the resampled
data sets, and so the \OIII\ line is more compact than the \Hb\ line
with 98.1\% confidence.

We also measure the spatially resolved $f(\OIII)/f(\Hb)$ ratio and
$(U-B)$ in the inner and outer regions of the stacked spectrum.  We
designate the ``inner region'' as the central $3 \times 3$ pixel box
($0\farcs39 \times 0\farcs39$, or $3.2 \times 3.2$~kpc at $z=1.86$) in
the cross-dispersion direction, and the ``outer region'' as the two
boxes 3 pixels wide from pixel numbers 4 through 6
($0\farcs39-0\farcs78$) above and below the center in the
cross-dispersion direction.  Line ratios are calculated for both the
inner and outer regions, using the same method of continuum
subtraction and Gaussian fitting described above.

Color gradients were calculated using the CANDELS WFC3 $Y$, $J$, and
$H$ imaging \citep{koe11}.  The WFC3 images are drizzled to a finer
resolution (0\farcs06) than the WFC3 spectroscopy, and so we adopt
$0\farcs36$ (6 pixel circular diameter) as the inner region and
$0\farcs36-0\farcs72$ (circular annulus from 6 to 12 pixels) as the
outer region.  Rather than stacking, we measure aperture $Y$, $J$, and
$H$ for each individual galaxy.  We then treat the median $Y-J$ (for
low redshift $z<1.8$) or $J-H$ (for $z>1.8$) difference from the inner
($<0\farcs36$) to outer ($0\farcs36-0\farcs72$) aperture as the change
in $(U-B)_{\rm rest}$ over the same aperture.

\begin{figure}
\scalebox{1.2}
{\plotone{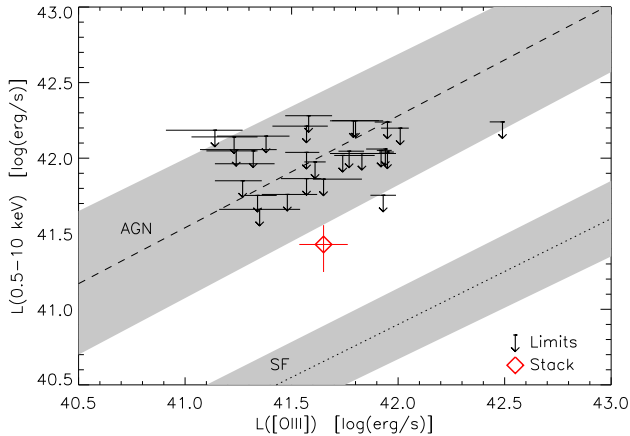}}
\figcaption{X-ray luminosity with \OIII\ luminosity for the 28 $z \sim
  2$ galaxies from the WFC3 G141 data.
%%   Only one object is detected in the 4~Ms {\it Chandra} data, shown by the
%%   blue circle.
  Since all of our galaxies are undetected in X-rays, we show them as
  X-ray upper limits.  The red diamond shows the stacked X-ray data
  (and mean $L_{\rm [OIII]}$) for the sample.  The dashed line shows
  the typical relation between $L_X$ and $L_{\rm [OIII]}$ for Type 2
  AGNs \citep{lam11} and the dotted line shows the typical relation
  for star-forming galaxies \citep{rov09}, both at $z \sim 0$, with
  the $1\sigma$ scatter about each relation indicated by the gray
  shaded area.  The position of the stacked X-ray data in $L_X-L_{\rm
  [OIII]}$ space suggests that an AGN is present in at least some of
  the galaxies.
\label{fig:xray}}
\end{figure}

As Figure \ref{fig:bpt} shows, the stacked inner region has marginally
redder colors and slightly higher line ratios than the outer region.
In other words, the inner region is more like an AGN than the outer
region.  However the large errors (particularly on the outer ratio)
mean that this is significant only at the $1.0\sigma$ level (67\%).
This is probably because the $0\farcs39$ size of the inner region
corresponds to the inner 3.2~kpc of the galaxy (at the median redshift
$z=1.86$), and so includes much more than the typical $\sim$1~kpc
narrow line region (NLR) of an AGN.  The marginal difference between
inner and outer line ratios is merely suggestive, but is reinforced by
the far more significant (98.1\% confidence) result that the \OIII\
profile is more compact than \Hb.

\subsection{Stacked {\it Chandra} Data}

In Figure \ref{fig:xray} we show the full-band (observed 0.5-10~keV)
X-ray luminosity upper limits with the \OIII$\lambda5007$ emission
line luminosity.  The dashed line shows the typical $L_X-L_{\rm
  [OIII]}$ relation for \OIII-selected narrow-line AGN \citep{lam11},
with associated $1\sigma$ scatter of $\sim$0.5~dex indicated by the
gray shaded area, and the dotted line shows the typical $L_X-L_{\rm
  [OIII]}$ relation for star-forming galaxies \citep{rov09}, with the
gray shaded area similarly indicating the $1\sigma$ scatter of
$\sim$0.3~dex.
%% The detected source, shown by the blue circle, has an high X-ray
%% luminosity ($1.5 \times 10^{42}$~erg~s$^{-1}$) indicative of a weak
%% AGN, and its position in $L_X-L_{\rm [OIII]}$ space suggests that it
%% has some combination of star formation and AGN activity contributing
%% to its \OIII\ emission.

We stack the X-ray data for the undetected sources to learn whether
the population has $L_X-L_{\rm [OIII]}$ dominated by black hole
activity or star formation.  The stacked data is marginally detected
($2.9\sigma$ significant) with a flux of $f_{\rm 0.5-10 keV}=1.1
\times 10^{-17}$~erg~s$^{-1}$~cm$^{-2}$.  At the median redshift of
our sample, $z=1.86$, this translates to an X-ray luminosity of
$L_{\rm 0.5-10 keV} = 2.7 \times 10^{41}$~erg~s$^{-1}$.  Because the
stack is only marginally detected in the full band, we cannot comment
on its hardness.  Unless the typical $L_X-L_{\rm [OIII]}$ relation for
star formation changes with redshift, the position of the stacked data
is intermediate between the loci of AGN and star formation.  This
suggests that at least some of the galaxies have an AGN contributing
to the \OIII\ and X-ray emission.

\section{Nature of $z \sim 2$ Emission-Line Galaxies}

The stacked set of our $z \sim 2$ galaxies shows a more compact
\OIII\ profile than the \Hb\ emission line.  This is the reverse of
what is expected for typical galaxy metallicity gradients
\citep{vanzee98}, but is expected if an AGN in the galaxy's center
contributes to the line emission.  We find a marginally higher
\OIII/\Hb\ in the center of the stacked data, similar to the
higher-confidence result of \cite{wri10}, who found a steeper
\OIII/\Hb\ ratio in the center of one $z=1.6$ galaxy.  In addition the
stacked {\it Chandra} data suggests an AGN component in at least some
objects, since the position of the stacked data in $L_X-L_{\rm
  [OIII]}$ is intermediate between the lines of star formation and
black hole activity.  For these reasons we suggest that some of these
$z \sim 2$ emission-line galaxies, and those with similar line ratios
and colors from other authors \citep{shap05b,erb06,kri07}, probably
contain weak AGN in their centers.

Because we rely on stacked data to reveal an AGN, it is difficult to
know the frequency of nuclear activity in these galaxies.  In the
local universe, AGN activity is known to correlate with SFR
\citep[e.g.][]{shi09}, so AGN activity might be expected in our
galaxies given their bright emission lines and presumably high star
formation rates.  While some authors have argued that AGN are rare in
local galaxies with $M_*<10^{10}M_{\odot}$ \citep{kau03}, recent work
suggests instead that AGN incidence is constant with host mass at
$z<1$ \citep{air11}.  Our sample has a median mass of
$M_*=10^{9.5}M_{\odot}$, and so unless a few massive objects are
dominating the stacked X-ray and grism spectrum, this suggests that
AGN are common in low-mass star-forming galaxies at $z \sim 2$.

\acknowledgements

JRT, and the other authors at UC Santa Cruz, acknowledge support from
NASA HST grant GO 12060.10-A, Chandra grant G08-9129A, and NSF grant
AST- 0808133.  We thank Stephanie LaMassa for access to her data, and
Shelley Wright for helpful discussions.  We additionally thank the
anonymous referee for helpful comments which improved the quality of
the manuscript.  We thank A. Riess (P.I.) and his SN team for
acquiring the grism data and acknowledge support from NASA HST grant
GO 12099.

\end{document}